\documentclass[preprint, a4paper, 3p, 11pt]{elsarticle}

\usepackage{graphicx}
\usepackage{amssymb}
\usepackage{amsmath}
\usepackage[section]{placeins}

\setcitestyle{authoryear}
\setcitestyle{round}
\setcitestyle{semicolon}

\usepackage{lineno}
\usepackage{url}
\usepackage[]{algorithm2e}
\usepackage{setspace}
\usepackage{braket}
\usepackage{mathrsfs}

\usepackage[dvipsnames]{xcolor}
\makeatletter
\def\ps@pprintTitle{%
 \let\@oddhead\@empty
 \let\@evenhead\@empty
 \def\@oddfoot{}%
 \let\@evenfoot\@oddfoot}
\makeatother

\journal{European Journal of Operational Research}

\begin{document}

%%%% DELETE BEFORE SUBMISSION %%%%
\let\today\relax
%%%%%%%%%%%%%%%%%%%%%%%%%%%%%%%%%%

\begin{frontmatter}

% Paper title
\title{Permanent and transitory crime risk in variable-density hot spot analysis}

% Author names and affiliations
\author[first,second,third]{Ben Moews\corref{corresponding}}
\ead{ben.moews@ed.ac.uk}
\cortext[corresponding]{Corresponding author}

\address[first]{Business School, University of Edinburgh, 29 Buccleugh Pl, Edinburgh, EH8 9JS, UK}
\address[second]{Centre for Statistics, University of Edinburgh, Peter Guthrie Tait Rd, Edinburgh, EH9 3FD, UK}
\address[third]{Scottish Centre for Crime and Justice Research, 63 Gibson St, Glasgow, G12 8LR, UK}

\date{}

% Abstract
\begin{abstract}
Crime prevention measures, aiming for the effective and efficient spending of public resources, rely on the empirical analysis of spatial and temporal data for public safety outcomes. We perform a variable-density cluster analysis on crime incident reports in the City of Chicago for the years 2001--2022 to investigate changes in crime share composition for hot spots of different densities. Contributing to and going beyond the existing wealth of research on criminological applications in the operational research literature, we study the evolution of crime type shares in clusters  over the course of two decades and demonstrate particularly notable impacts of the COVID-19 pandemic and its associated social contact avoidance measures, as well as a dependence of these effects on the primary function of city areas. Our results also indicate differences in the relative difficulty to address specific crime types, and an analysis of spatial autocorrelations further shows variations in incident uniformity between clusters and outlier areas at different distance radii. We discuss our findings in the context of the interplay between operational research and criminal justice, the practice of hot spot policing and public safety optimization, and the factors contributing to, and challenges and risks due to, data biases as an often neglected factor in criminological applications. 
\end{abstract}


\end{frontmatter}

%%%% CHANGE BEFORE SUBMISSION %%%%
\nolinenumbers %linenumbers
%%%%%%%%%%%%%%%%%%%%%%%%%%%%%%%%%%

\onehalfspacing

% Paper main body
\section{Introduction}
\label{sec:introduction}

Having arguably started with an essay by \citet{Guerry1833}, the field of mathematical methods in criminology is an important part of criminal justice endeavors in modern society, and scholars have taken a renewed interest in data collection and standardization in the early twentieth century, with the emerging field of modern operational research soon picking up the baton \citep{Frankel1928, Wilkins1954}. These efforts quickly enabled the study of correlations between population densities and crime as well as variations in incident occurrences between cities \citep[see, for example,][]{Watts1931, Ogburn1935}, and are the legacy of today's literature on intra-city crime analysis.

In this context, ``hot spots'' refer to the clustering of crime in subregions of an area of interest for policing, with studies demonstrating the efficacy for crime reduction \citep[see][for a systematic review and meta-analysis]{Braga2014}. \citet{Weisburd2018} report that 90.8\% of U.S. police departments have adopted hot spot policing as an approach, with three quarters of the nation's agencies having done so through a formal policy. As a result, research on this topic is both timely and pressing, as decision-making in this area, when not steered by empirical data and sound quantitative evaluation, invites the risk of biases and inequality through the targeting of specific demographics \citep{Maltz1975, Wheeler2020}.

One particular challenge are high-density peaks leading to narrow clusters and a focus on population density instead of structurally similar areas. In our analysis, we will see that one such example is the Chicago Loop, the main section of Downtown Chicago containing large parts of the city's entertainment and restaurant scene, and one of North America's largest business districts. Our study follows calls for the broader inclusion of non-epicenter hot spot areas, for example by \citet{Eck2005}, and acts upon recent demands for research on varying-density clustering in crime analyses \citep{Xie2019, Cesario2022}.

The outbreak of the recent COVID-19 pandemic has led to lockdowns and social distancing measures across the globe, including the City of Chicago \citep{ScannellBryan2021}. An impact on criminal activity is a natural conjecture, and prior works demonstrate an effect on the composition of crimes, as well as differences in that change between locations, although existing analyses are limited to the resolution of community areas without targeting crime clustering \citep{Boman2020, Campedelli2020, Kim2022}. 

This case study, which focuses on Part I crimes, makes several new contributions to the operational research literature on hot spot analysis. We present the first adaption and application of recent advances in continuous distance matrix rescaling in geospatial clustering to criminological data, and provide a large-scale analysis of more than two decades worth of crime incident reports for the City of Chicago in the process. Our experiments include the spatio-temporal investigation of crime shares relative to cluster number densities across years, and showcase the changing composition of different hot spots within the city boundaries. Aside from researchers working in the same area, our findings are of interest to practitioners in criminal justice and can inform both crime prevention measures through policing and decision-makers in public policy.

Our results on crime shares relative to the primary function of city areas also highlight the impact of the COVID-19 pandemic, and are the first demonstration of these effects in the field of hot spot analysis. The experiments performed for this paper also include a methodology transfer of spatial two-point autocorrelations for clumping effects in cosmology, providing an example or the interdisciplinary applicability of methods developed in other fields. The associated findings show that crime hot spots identified with our approach exhibit additional granular features compared to the rest of the city, providing further evidence for the importance of diffusing policing outside of unchanging density epicenters \citep{Saksena1979, Camacho-Collados2015}. Lastly, we discuss the risk of data biases with regard to discrepancies between reported incidents and underlying crime rates, as well as the effect of police-community relations on accurate crime statistics.

The remainder of this paper is structured as follows. Section~\ref{sec:background_and_methodology} provides an overview of hot spot policing, as well as descriptions of our methodology and data. Section~\ref{sec:empirical_analysis_and_results} covers our experiments on spatial clustering, temporal intra-cluster changes, and data uniformity. Section~\ref{sec:discussion_and_study_limitations} discusses our results and their impact, avenues for follow-up research, and limitations of both the methodology and potential data biases, with a particular focus on the last point as a topic that often receives only a short mention in the related literature. Lastly, we provide our conclusions in Section~\ref{sec:conclusion}.

\section{Background and methodology}
\label{sec:background_and_methodology}

\subsection{Hot spot analysis and patrolling}
\label{subsec:hot_spot_analysis_and_patrolling}

The notable clustering effects of crime are well-document for urban environments and have been coined the ``law of crime concentration'' accordingly \citep{Weisburd2015}. When policing such hot spots, the question of geographical constancy is important for planning, and research in geospatial analysis has demonstrated the persistence of crime clusters over time, as well as the need to allocate public resources efficiently to police hot spots \citep{He2017, Leigh2019}. 

For the purpose of identifying hot spots, as well as for information sharing, computational methods relying on empirical data are now the nearly exclusive approach used in police departments across the United States, although different constraints apply to rural patrol settings \citep{Birge1989, Redmond2002, Reaves2007}. When employing these methods, data reliability and the risk of biases that can affect the latter enter the picture, and we will dedicate part of the discussion in Section~\ref{sec:discussion_and_study_limitations} to this ongoing challenge for researchers, as the application of quantitative methods in this area is subject to ethical considerations \citep{Kleijnen2001}.

The majority of the criminology literature, together with approaches employed in practice, focuses on kernel density estimation and parametric partitioning methods as an established and easily accessible way to search for hot spots \citep{Novak2016, Weisburd2018}. The operational research and statistics communities, on the other hand, have spread their efforts across a wider area, with varying foci depending on the method in question, and \citet{Blumstein2007} stresses the existing role of operational research in terms of a missionary function for quantitative methods in the criminal justice system in particular \citep{Larson2002}.

Scan statistics have enjoyed some popularity in this field, for example using simulated annealing \citep{Duczmal2004}. One issue with these approaches are potential biases regarding the morphological structure of clusters, often employing circular scanning windows for their geometric compactness as pointed out by \citet{Kulldorf2006}, although methods such as support vector scan statistics have been developed to address this problem~\citep{Fitzpatrick2021}. \citet{Grubesic2006} also point out the risk of polygonal units capturing incorrect relationships when using standard adjacency metrics.

Another active area of study in the literature is centered on self-excited point processes, which is more concerned with the predictive capacities of models fitted to crime data as an underlying probability distribution  \citep[see, for example,][]{Park2021}. More recently, other methods have been the subject of renewed interest, for example Bayesian hierarchical modeling to encourage spatial smoothness within clusters at the neighborhood level, exemplifying the continuous methodological innovation in this area of application \citep{Balocchi2023}. An up-to-date literature review on applications to patrolling in operational research can be found in \citet{Samanta2022}.

These developments address earlier criticisms in the field of criminology that density plotting by itself is insufficient, as argued by \citep{Eck1997}, and cluster analysis for understanding the dynamics of crime in high-impact regions has been proposed to be important to regional planners, policy makers, and policing agencies, as crime prevention strategies require suitable patrol coverage and response times \citep{Olson1975, Bammi1976, Murray2001, Adler2014}. When doing so, however, both the crime type composition of persistent hot spots, as shown by \citet{He2017}, and the changes in the latter over time are important.

Given the disruptive nature of the COVID-19 pandemic, such changes naturally also emerged in this context. Having been declared a pandemic in early 2020, this led to lockdowns and social distancing measures across the world, including in Chicago as the source of our dataset \citep{ScannellBryan2021, Eryarsoy2023}. The approximate delineation to the preceding year also allows for an easy incorporation into annual analyses. Effects on crime incidents have been investigated since shortly after the outbreak, noting an overall decrease in crime which is driven by decreases in minor offenses committed in peer groups. Conversely, crimes without co-offenders such as homicide and domestic violence are reported as having either remained constant or increased in the United States, which is also reflected in the literature on homicide in particular being resistant to environmental changes \citep{Anderson1996, Boman2020}.

Studies on these effects include Chicago in particular, with \citet{Campedelli2020} using a Bayesian structural time series approach for a subset of offenses to show that crime trend changes vary between communities and types of crimes. This is further corroborated by \citet{Kim2022}, who report on a causal relationship between lockdown measures and a decrease in battery and sexual assault, as well as an increase in homicides, with the same dependence on location and crime type. These multifaceted changes make an investigation of hot spots as a core tool of urban policing a natural follow-up in the tradition of operational research informing this application area.

\subsection{Clustering and density rescaling}
\label{subsec:clustering_and_density_rescaling}

\citet{Murray2014} investigate spatial clustering techniques in a range of applied areas, including crime analysis, and stress the importance of contiguous and irregularly-shaped clusters to avoid spatial biases. These sentiments are echoed in related research, with a focus on the relevance of discontinuities between areas for subsequent local analyses and the lack of shape constraints due to varied urban layouts \citep{Yin2021}.

As practitioners often emphasize the importance of hot spot epicenters, leading to the widespread application of ``cops on dots'' patrolling, the remainder of cluster areas are often underpatrolled \citep{Eck2005}. Following these assessments, our chosen method alleviates these issues within a framework familiar to the criminology literature by using an extended version of the density-based spatial clustering of applications with noise (DBSCAN) algorithm. First introduced by \citet{Ester1996}, it has found recent success in the application to criminological data \citep[see, for example,][]{Chen2020, Robertson2022}. In this approach,
\begin{equation}
N_{\epsilon}(x) = \{ y \in D \ | \ \delta(x,y)\leq \epsilon \}
\label{eq:dbscan_neighborhood}
\end{equation}
defines the $\epsilon$-neighborhood for the distance $\delta(x,y)$ between two points from a set $D \in \mathbb{R}^2$, with $\epsilon$ as the maximum contiguity radius. The algorithm then employs three types of reachability; a point is classified as directly density-reachable for
\begin{equation}
y \in N_{\epsilon}(x), \ \mathrm{with} \ |N_{\epsilon}(x)| \geq \kappa,
\label{eq:dbscan_ddr}
\end{equation}
for a minimum number of cluster members $\kappa$ as the core point condition, thus favoring cluster identification in high-density regions. Next, density-reachability between a point set $\{x, y\}$ for a chain $\{p_1, p_2, \dots, p_t\}$, with $p_1 = x$ and $ p_t = y$, is given in the case of
\begin{equation} 
\begin{split}
&\omega_\mathrm{r}(p_1, p_t) := \forall p_i \in \{p_1, p_2, \dots, p_{t-1}\} :\\
&p_{i+1} \in N_\epsilon(p_i),
\end{split}
\label{eq:dbscan_dr}
\end{equation}
introducing a linkage condition to form irregularly-shaped clusters. The third type defines density-connected points, which requires a third point $z$ in terms of a second chain $\{ q_1, q_2, \dots q_t \}$, redefining $y = q_1$ and setting $z = p_{t+1} = q_{t+1}$,
\begin{equation} 
\begin{split}
&\forall p_i \in \{ p_1, p_2, ... p_{t} \} \forall q_i \in \{ q_1, q_2, \dots q_{t} \} :\\
&p_{i+1} \in N_\epsilon(p_1) \land q_{i+1} \in N_\epsilon(q_1), 
\end{split}
\label{eq:dbscan_dc}
\end{equation}
allowing for linkage through a  density-reachable point not fulfilling the core point condition in Equation~\ref{eq:dbscan_ddr}, in which case direct density-reachability ceases to be symmetric. This reachability notion is both symmetric and, for density-reachable points, reflexive. We can alternatively write density-connectivity as $\omega_\mathrm{c}(x, y)$ for both $\omega_\mathrm{r}(x, z) \ \mathrm{and} \ \omega_\mathrm{r}(y, z)$, which is transitive but not symmetric, and test for any $\emptyset \neq C \subseteq D$ as a cluster through
\begin{equation}
x \in C \land \omega_\mathrm{r}(x, y) \Rightarrow y \in C
\label{eq:dbscan_cluster_1}
\end{equation}
as the first of two criteria for cluster building, with the second given by
\begin{equation}
\forall x, y \in C : \omega_\mathrm{c}(x, y).
\label{eq:dbscan_cluster_2}
\end{equation}
Lastly, the noise set $N \subseteq B$ contains points for which, given clusters $\{ C_1, C_2, \dots, C_n \} \subseteq D$,
\begin{equation}
N := \{ x \in D \ | \ \forall i : x \notin C_i \}, 
\label{eq:dbscan_noise}
\end{equation}
which is simply to say that all points not part of a cluster as per Equations~\ref{eq:dbscan_cluster_1}--\ref{eq:dbscan_cluster_2} are considered outliers. Now that we have defined the basic functionality of DBSCAN, which enables the identification of contiguous and irregularly-shaped clusters, we need to address the previously mentioned concern regarding the identification of clusters too narrow to be useful. This is a broader challenge in the literature, as small regions of high density can make the identification of clusters in lower-density areas challenging in practical setttings \citep{Pei2009}. 

\citet{Gieschen2022} introduce, for spatio-temporal risk analysis, the continuous density-based rescaling of the distance matrix to alleviate this problem, although adaptions to the given problem are necessary for our resulting implementation of the method. By first applying a kernel density estimation to a set of spatial coordinates $D \in \mathbb{R}^2$ with $n := | D |$,
\begin{equation}
\hat{f} (x, \beta) = \frac{1}{n} \sum\limits_{i = 1}^{n} \frac{1}{(2 \pi \beta^2)^{\frac{1}{2}}} \exp\left( - \frac{|| x - D_i ||^2}{2 \beta^2} \right),
\label{eq:kde}
\end{equation}
we maintain the use of a Gaussian kernel standard deviation and optimization of $\beta$ following \citet{Scott1992}, which satisfies integration to unity and non-negativity everywhere. Let $\bar{g}$ be the arithmetic density average for the set  $\{ g_1, g_2, \dots, g_n \}$ of densities for points in $D$, then
\begin{equation}
\varphi(g_i) = \frac{2}{1 + e^{- k (g_i - \overline{g})}}
\label{eq:logistic}
\end{equation}
provides us with a logistic function on the interval (0, 2) centered on one. When we use this function as a multiplicator to rescale a distance matrix based on each point's density value, this ensures that no rescaling takes place for exactly average densities, whereas points closer to the average are subject to larger differences in rescaling. 

Similarly, the lower limit of zero is only applied in cases in which points share the same values in both dimensions, in which case they already have the same distance and nothing changes. Given a distance between a pair of points $\delta(D_i, D_j)$, the rescaled value  $\hat{\delta}(D_i, D_j)$ is
\begin{equation}
\begin{split}
\hat{\delta}(D_i, D_j) &= \delta(D_i, D_j) \cdot \left( \frac{\varphi(g_i) + \varphi(g_j)}{2} \right)\\
&= \delta(D_i, D_j) \cdot \frac{\left( \frac{2}{1 + e^{- k (g_i - \overline{g})}} + \frac{2}{1 + e^{- k (g_j - \overline{g})}} \right)}{2}.
\end{split}
\label{eq:rescaled}
\end{equation}
While the above parts provide the approach for cluster detection, we will later also be interested in the pre-rescaling uniformity of clusters versus outliers, which is further analyzed in Section~\ref{subsec:distributions_and_clumping_of_crime} in the context of intra-cluster policing beyond epicenters.

While \citet{Gieschen2022} use the Meeus calculation of distances on an obloid approximation of Earth, popularized by \citet{Meeus1991} for astronomical applications, the main disadvantage of this formula is the often prohibitive computational cost for large-scale applications. For our study, we adopt the Euclidean distance into our implementation of the above method, as geographical features play a considerably larger role than the planet's curvature at the localized city level.

\subsection{Uniformity and autocorrelations}
\label{subsec:uniformity_and_autocorrelations}

Rooted in seminal work by \citet{Ripley1976} \citep[see also][]{Ripley1977}, ``Ripley's alphabet'' is a set of functions for distance-based investigations of clustering in point processes \citep{Baddeley2015}. We define the empty space distance, for a random location $u \in \mathbb{R}^2$, as
\begin{equation}
\delta_\mathrm{e}(u, D) = \min \{ || u - D_i || : D_i \in D \}.
\label{eq:empty_space}
\end{equation}
For a given stationary point process $X$, the empirical distribution function of the observed empty space distances can, for a grid $\{ u_1, u_2, \dots, u_m \}$, be written as
\begin{equation}
\hat{F}(r) = \frac{1}{m} \sum_{i=1}^m 1 \{ \delta_\mathrm{e}(u_i, D) \leq r \}.
\label{eq:f_empirical}
\end{equation}
Taking a homogeneous Poisson process with intensity $\lambda$ as the benchmark, this becomes
\begin{equation}
F_\mathrm{p}(r) = 1 - \exp(-\lambda \pi r^2),
\label{eq:f_poisson}
\end{equation}
allowing us to plot the observed estimate versus the Poisson process, with $\hat{F}(r) < F_\mathrm{p}(r)$ indicating clustering due to empty space distances that are larger than would be expected from the benchmark. Similarly, we can write nearest-neighbor distances as
\begin{equation}
\delta_\mathrm{n}(D_i) = \min_{i \neq j} || D_i - D_j ||,
\label{eq:nearest_neighbor}
\end{equation}
with an empirical distribution function for the observed nearest-neighbor distances
\begin{equation}
\hat{G}(r) = \frac{1}{n} \sum_{i=1}^n 1 \{ \delta_\mathrm{n}(D_i) \leq r\}.
\label{eq:g_empirical}
\end{equation}
Analogous to the empty-space functions above, the formulation for the homogeneous Poisson process $G_\mathrm{p}(r)$ remains the same as in Equation~\ref{eq:f_poisson}. In contrast to the empty-space functions, however, clustering is indicated for $\hat{G}(r) > G_\mathrm{p}(r)$ through nearest-neighbor distances shorter than would be expected from a Poisson process. While these functions will serve an important role to gain an overview over the dataset used in this paper, our subsequent comparison between hot spots and the non-clustered remainder in terms of spatial uniformity requires a method that allows us to more easily view and compare results for multiple clusters at once, and to calculate confidence intervals to gauge the robustness of detections in our subsequent applications.

In the spirit of interdisciplinarity, which is central to the role of operational research, we shift our gaze upward to the field of astrostatistics, where the spatial two-point autocorrelation function is an established core constituent of modern cosmology, and is often referred to as simply the correlation function due to its ubiquity \citep{Verde2010}. It is commonly used for distance measurements averaged over randomly-sampled galaxies, stating the excess probability for identifying a pair of galaxies for a given distance when compared to uniform probability. As such, it is another way to measure clustering effects for different distance scales. It can be written as 
\begin{equation}
\xi_2(D_i , D_j) = \braket {\delta(D_i) \delta(D_j)}
\label{eq:two_point}
\end{equation}
for the calculation of the covariance for a given field $\rho(D)$ measured at two points, with 
\begin{equation}
\delta(D) = \frac{\rho(D) - \bar{\rho}}{\bar{\rho}}
\label{eq:overdensity}
\end{equation}
as a unitless overdensity measure for the density field average $\bar{\rho}$. The underlying assumptions are isotropy and homogeneity, meaning uniformity in all directions and at every point, which is the case for uniform distributions in $\mathbb{R}^2$ as used in our case. In practice, this is mostly estimated through counts of observed data pairs $\vartheta_{\mathrm{DD}}$ and counts of data pairs $\vartheta_{\mathrm{RR}}$  for a generated Poissonian dataset $R$. The most basic formulation, for a given distance $r$ and number densities $\nu(D)$ and $\nu(R)$ for the observed and generated dataset, respectively, is
\begin{equation}
\hat{\xi} (r) = \frac{\nu(R) \vartheta_{\mathrm{DD}}(r)}{\nu(D) \vartheta_{\mathrm{RR}}(r)} - 1,
\label{eq:2pCF_standard}
\end{equation}
or, more accurately, through the Landy-Szalay estimator by \citep{Landy1993}, 
\begin{equation}
\hat{\xi}_\mathrm{LS} (r) = \frac{\vartheta_{\mathrm{DD}}(r) \left( \frac{\nu(R)}{\nu(D)} \right)^2 - 2 \vartheta_{\mathrm{DR}}(r) \left( \frac{\nu(R)}{\nu(D)} \right) + \vartheta_{\mathrm{RR}}(r)}{\vartheta_{\mathrm{RR}}(r)},
\label{eq:2pCF_ls}
\end{equation}
with counts between the observed and randomly-generated data denoted as $\vartheta_{\mathrm{DR}}$. This is especially useful due to the variance being Poissonian in the limit of weak clustering and $1 / |D|$ otherwise, while points near the edge of a sample space are contributing in a prorated fashion. 

In Section~\ref{subsec:distributions_and_clumping_of_crime}, this approach will help us to assess differences in uniformity between hot spots and the rest of the dataset. Operational research lives at the intersection of various areas of research, both in terms of other mathematical sciences and areas of domain application, and we wish to showcase the interdisciplinary utility of methods that are often developed and, after their dissemination within the respective research communities, constrained to separate fields. 

\subsection{Data access and urban policing}
\label{subsec:data_access_and_urban_policing}

The availability of reliable and standardized crime statistics is a crucial step for the quantitative analysis of crime, and its importance has long been stressed by researchers working in this area \citep{Beattie1959}. Our case study is centered on Part I crimes as defined by the Uniform Crime Reports (UCR) published by the U.S. Federal Bureau of Investigation. The Chicago Data Portal\footnote{\url{https://data.cityofchicago.org}}, the open-access outlet of the Citizen Law Enforcement Analysis and Reporting (CLEAR) system of the Chicago Police departments, provides access to suitable reported crime incidents from 2001 onward, except for homicides with data for separate victims.

This dataset features block-level GPS coordinates for incidents and is updated daily, with a one-week delay of dates covered, making it a prime target for urban geospatial analysis. Due to this availability, the database is frequently used for the study of urban crime, for example in the context of police patrol planning \citep{Chen2017, Moews2021}. We obtain the complete dataset from 2001 to the end of 2022, covering 22 years of geocoded Part I crime reports for the area of the City of Chicago in 7,710,973 entries, with each entry featuring data for the year, primary crime type, and latitude and longitude coordinates.

\begin{table*}[!htb]
\caption{Part I crime incident reports for the City of Chicago from 2001--2022, as available from the Chicago Data Portal. The table lists, for each year, the reports of robbery, theft, burglary, motor vehicle theft, assault, criminal sexual assault, arson, and homicide, with the first two letters of each crime type indicated in the column headers.}
\begin{footnotesize}
\begin{center}
\begin{tabular}{lllllllll}
\hline
Year & RO & TH & BU & MO & AS & CR & AR & HO \\
\hline
2001 & 18292 & 98447 & 25943 & 27282 & 31260 & 1763 & 1005 & 666 \\
2002 & 17740 & 95363 & 25221 & 23255 & 30733 & 1701 & 978 & 658 \\
2003 & 17235 & 97804 & 25010 & 22676 & 29292 & 1534 & 953 & 603 \\
2004 & 15951 & 94642 & 24520 & 22747 & 28792 & 1467 & 774 & 455 \\
2005 & 15988 & 84304 & 25413 & 22384 & 26965 & 1422 & 688 & 453 \\
2006 & 15943 & 85233 & 24304 & 21785 & 25929 & 1370 & 726 & 476 \\
2007 & 15445 & 84600 & 24838 & 18553 & 26305 & 1460 & 710 & 448 \\
2008 & 16590 & 86406 & 26012 & 18626 & 25273 & 1413 & 643 & 514 \\
2009 & 15848 & 79305 & 26495 & 15313 & 22616 & 1305 & 612 & 514 \\
2010 & 14272 & 76739 & 26421 & 19026 & 21534 & 1336 & 522 & 438 \\
2011 & 13977 & 75123 & 26616 & 19384 & 20406 & 1451 & 504 & 438 \\
2012 & 13483 & 75444 & 22840 & 16488 & 19897 & 1392 & 469 & 515 \\
2013 & 11819 & 71501 & 17893 & 12576 & 17969 & 1252 & 364 & 431 \\
2014 & 9795 & 61458 & 14562 & 9895 & 16889 & 1269 & 396 & 429 \\
2015 & 9632 & 56696 & 13103 & 10003 & 16992 & 1272 & 453 & 502 \\
2016 & 11953 & 61038 & 14280 & 11270 & 18720 & 1495 & 515 & 790 \\
2017 & 11871 & 63585 & 12946 & 11339 & 19251 & 1530 & 444 & 676 \\
2018 & 9677 & 64024 & 11690 & 9934 & 20342 & 1566 & 373 & 601 \\
2019 & 7990 & 61680 & 9635 & 8963 & 20601 & 1582 & 375 & 508 \\
2020 & 7848 & 40223 & 8704 & 9893 & 18207 & 1145 & 587 & 796 \\
2021 & 7899 & 39258 & 6605 & 10487 & 20254 & 1411 & 525 & 809 \\
2022 & 8959 & 53152 & 7532 & 21295 & 20699 & 1509 & 419 & 713 \\
\hline
\end{tabular}
\end{center}
\label{tab:table_1}
\end{footnotesize}
\end{table*}

When dealing with empirical data, missing values are a regular feature, and dropping rows with such omissions reduces our data by 85,695 entries. Fortunately, this only represents around $1.1\%$ and leaves us with 7,625,278 records, split into years and crime types in Table~\ref{tab:table_1}. Similarly, as is the case with many real-world datasets, we have to embrace some limitations. One of these is a small randomization factor for data privacy, although the latter ensures that locations stay within the same block\footnote{\url{https://chicagostudies.uchicago.edu/grid}} with one eighth of a mile on each side. 

Another limitation is the nature of crime incident reports, as crime reporting can be subject to differences in police-community relations as a potential source of bias effects \citep[see, for example,][]{Slocum2010, Xie2012, Kochel2018}. While this presents an unavoidable component of the data gathering process, we will spend part of the discussion in Section~\ref{sec:discussion_and_study_limitations} on these limitations to place our analysis in the proper context.

\begin{figure*}[!htb]
\includegraphics[width=\textwidth]{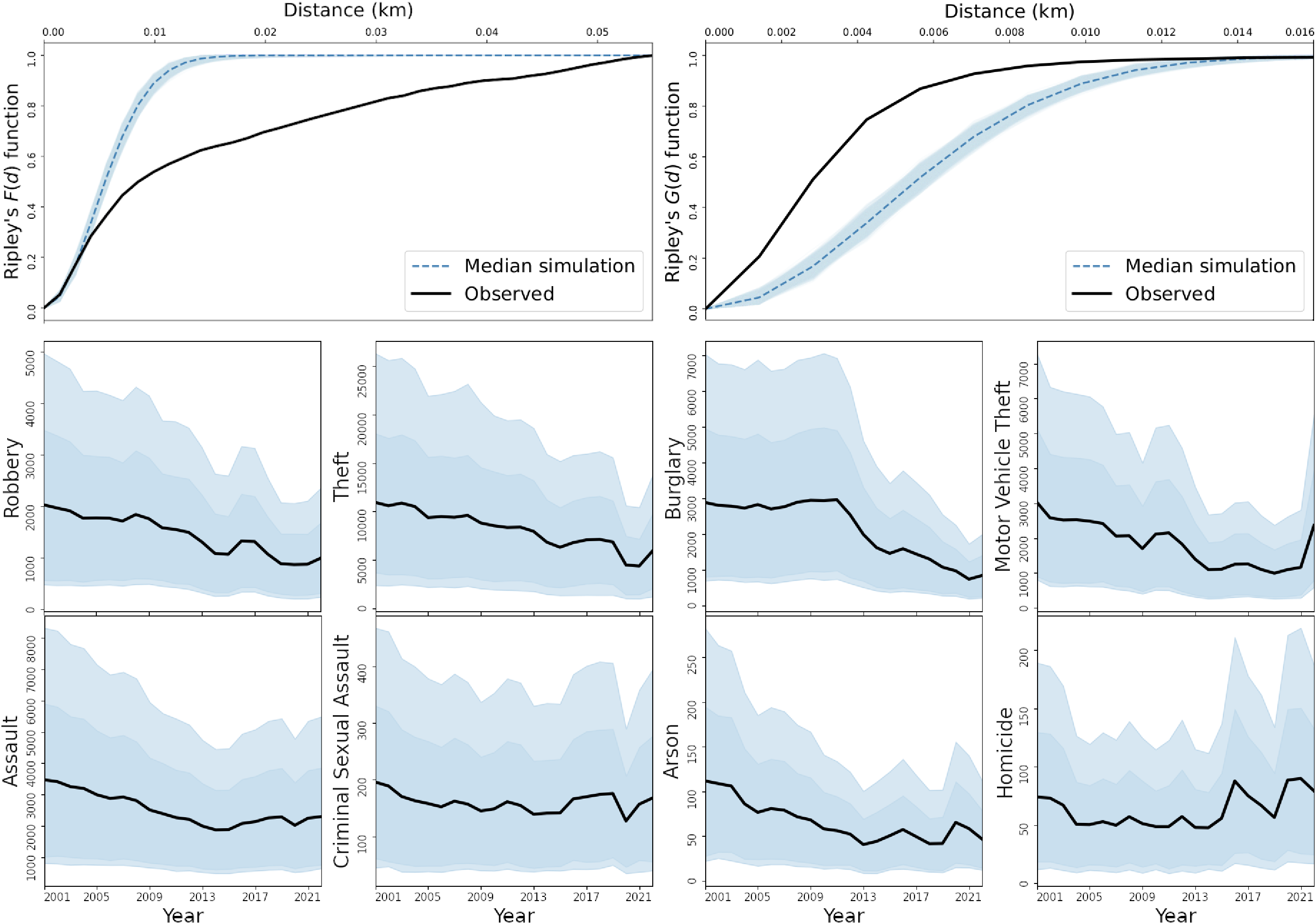}
\caption{Descriptions of uniformity for crime distributions and types for the City of Chicago from 2001--2022. The upper two panels show Ripley's $F(d)$ and $G(d)$ functions in black, for a given distance $d$, with the median of Poisson simulations as a dashed line and their 95\% confidence interval shaded. The lower eight panels depict the mean number of crimes per year across identified clusters in black, with darker and lighter shading indicating 95\% and 68\% confidence intervals for the different clusters, respectively.}
\label{fig:figure_1}
\end{figure*}

The upper part of Figure~\ref{fig:figure_1} shows Ripley's $F(d)$ and $G(d)$ functions as defined in Equations~\ref{eq:f_empirical} and~\ref{eq:g_empirical}, respectively, demonstrating clustering effects in our unprocessed dataset. For an overview separated by crime types, the lower part of the figure also shows the mean number of incidents, with confidence intervals for subsequently identified clusters to demonstrate the wide variability between the localities that motivate our study.

One point of note for researchers interested in applying these data, showcasing the importance to check for and resolve inconsistencies in datasets, is that both ``Criminal Sexual Assault'' and ``Crim Sexual Assault'' are used in the CLEAR system described in Section~\ref{subsec:data_access_and_urban_policing}. The latter, while only being used for a small fraction of incidents in earlier years, has exponentially overtaken the former during the last few years. These descriptors need to be aligned to avoid incorrect calculations.

\section{Empirical analysis and results}
\label{sec:empirical_analysis_and_results}

\subsection{Clustering and hot spot description}
\label{subsec:clustering_and_hot_spot_description}

In the first step, we need to assign all entries to identified hot spots. Due to the constraint of computationally expensive distance matrix calculations, we apply the variable-density cluster analysis described in Section~\ref{subsec:clustering_and_density_rescaling} on a sample following the methodology of \citet{Gieschen2022}. In contrast to the latter, our use of a locally sufficient distance metric leads to complete calculations within a few minutes. 76.60\% of entries are identified as outliers, which is in agreement with the goal to extract hot spots for a geographical area. We then assign the remainder of the dataset to clusters or outlier status using envelopes of ten meters around assigned samples, with a slightly lower outlier share of 70.29\% due to our envelope approach in close agreement.

Incident shares per cluster, as well as the sum, arithmetic mean, and standard deviation with respect to the years covered in our dataset are listed in Table~\ref{tab:table_2}, and reiterate the dominance of high-density peaks, which in this case refers to Cluster 1 due to its location in the Chicago Loop area previously mentioned in Section~\ref{sec:introduction} as a prime example of this urban feature.

\begin{table*}[!htb]
\caption{Statistical indices for outliers and identified crime clusters C1--C8 in the City of Chicago for 2001--2022. The table lists the share of overall crime, as well as the sum of crimes, arithmetic mean, and standard deviation over the years covered in the dataset.}
\begin{footnotesize}
\begin{center}
\begin{tabular}{lrrrrrrrrr}
\hline
& Outliers & C1 & C2 & C3 & C4 & C5 & C6 & C7 & C8 \\\hline
Share & 70.29\% & 8.09\% & 4.69\% & 1.89\% & 3.32\% & 4.34\% & 2.53\% & 2.19\% & 2.68\% \\
$\Sigma_\mathrm{years}$ & 2273191 & 261485 & 151579 & 61111 & 107214 & 140271 & 81852 & 70679 & 86593 \\
$\mu_\mathrm{years}$ & 103326.86 & 11885.68 & 6889.95 & 2777.77 & 4873.36 & 6375.95 & 3720.55 & 3212.68 & 3936.05 \\
$\sigma_\mathrm{years}$ & 26676.14 & 2177.97 & 1933.85 & 905.02 & 1500.64 & 1689.87 & 1181.97 & 818.66 & 665.48 \\
\hline
\end{tabular}
\end{center}
\end{footnotesize}
\label{tab:table_2}
\end{table*}

\begin{figure*}[!htb]
\includegraphics[width=\textwidth]{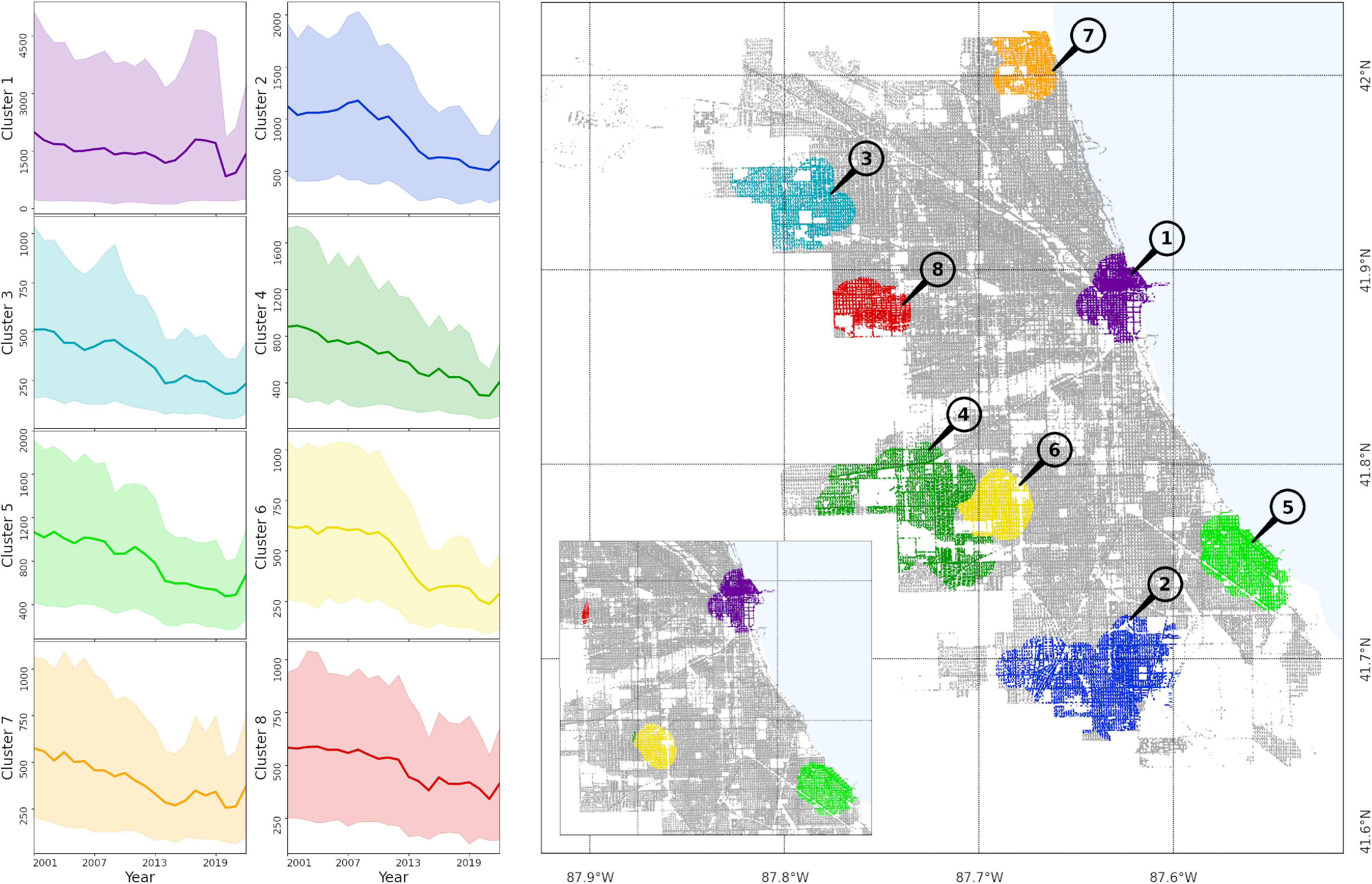}
\caption{Clustering of Part I crime reports for the City of Chicago. The eight left-hand panels show the mean number of reports for different crime types, with the shaded regions indicating the 95\% confidence intervals across types. The main right-hand panel depicts incidents falling within Clusters 1--8 with number flags, as well as non-cluster incidents in grey, with the subplot showing the agreement with a non-varying-density clustering.}
\label{fig:figure_2}
\end{figure*}

With an algorithm that does not require the setting of a cluster number as a parameter, we retrieve eight hot spots that are shown and numbered in the right part of Figure~\ref{fig:figure_2}. The agreement with the non-adjusted base algorithm as used on crime data before, for example by \citet{Chen2020} and \citet{Robertson2022}, is shown in the right-hand subplot to demonstrate the broad overlap with Clusters 1, 5, and 6. While a sliver of Cluster 8 is covered as well, the remaining clusters are not identified. As the dataset operates on a block level, The left part of Figure~\ref{fig:figure_2} shows the number of incidents per year and cluster, averaged over the Part I crime types represented in CLEAR data, to visualize the evolution of clusters over time.

The plots show that the well-documented overall decrease in crime, as outlined by \citet{Tcherni-Buzzeo2019}, is also present in the identified hot spots in Chicago. Narrowing confidence intervals reflect a decrease in the difference between crime types over time, although the potency of this effect varies across clusters and does not apply to all Part I crimes equally, as shown previously in Figure~\ref{fig:figure_1}. Robberies and burglaries, for example, are defined by a marked downward trend, while motor vehicle thefts experienced a recent surge, and criminal sexual assaults and homicides paint a picture of a constant presence level.

The right part of Figure~\ref{fig:figure_2} also highlights one result of the unavoidable block-level data analysis, with natural spaces and major traffic arteries being clearly visible in the collection of data points. One prominent example is the larger area around Lake Calumet in the south of the city, to the right-hand side of Cluster 2, which features multiple parks. Similarly, the same effect also disrupts clusters, with the most noticeable example being Cluster 4 with the Chicago Midway International Airport and Marquette Park, as well as Cluster 3 with Mount Olive Cemetery and Zion Gardens Cemetery, as well as multiple park areas. This part of our analysis demonstrates the direct impact of urban layouts on data collection efforts, which we will discuss further in Section~\ref{subsec:distributions_and_clumping_of_crime}.

\subsection{Fixed patterns and temporary shifts}
\label{subsec:fixed_patterns_and_temporary_shifts}

Next, we direct our attention to the evolution of separate Part I crimes over time, and split this analysis into the identified hot spots as well as the surrounding outliers. Scaling is required to enable a sensible comparison, as hot spots feature varying numbers of incidents. Similarly, given that the majority of entries are not part of hot spots, unscaled values would also make visualizations unusable, as their respective counts would drown out clustered subsets and pool the latter at the bottom of graphs, and related research often only visualizes separate crime types.

\begin{figure*}[!htb]
\includegraphics[width=\textwidth]{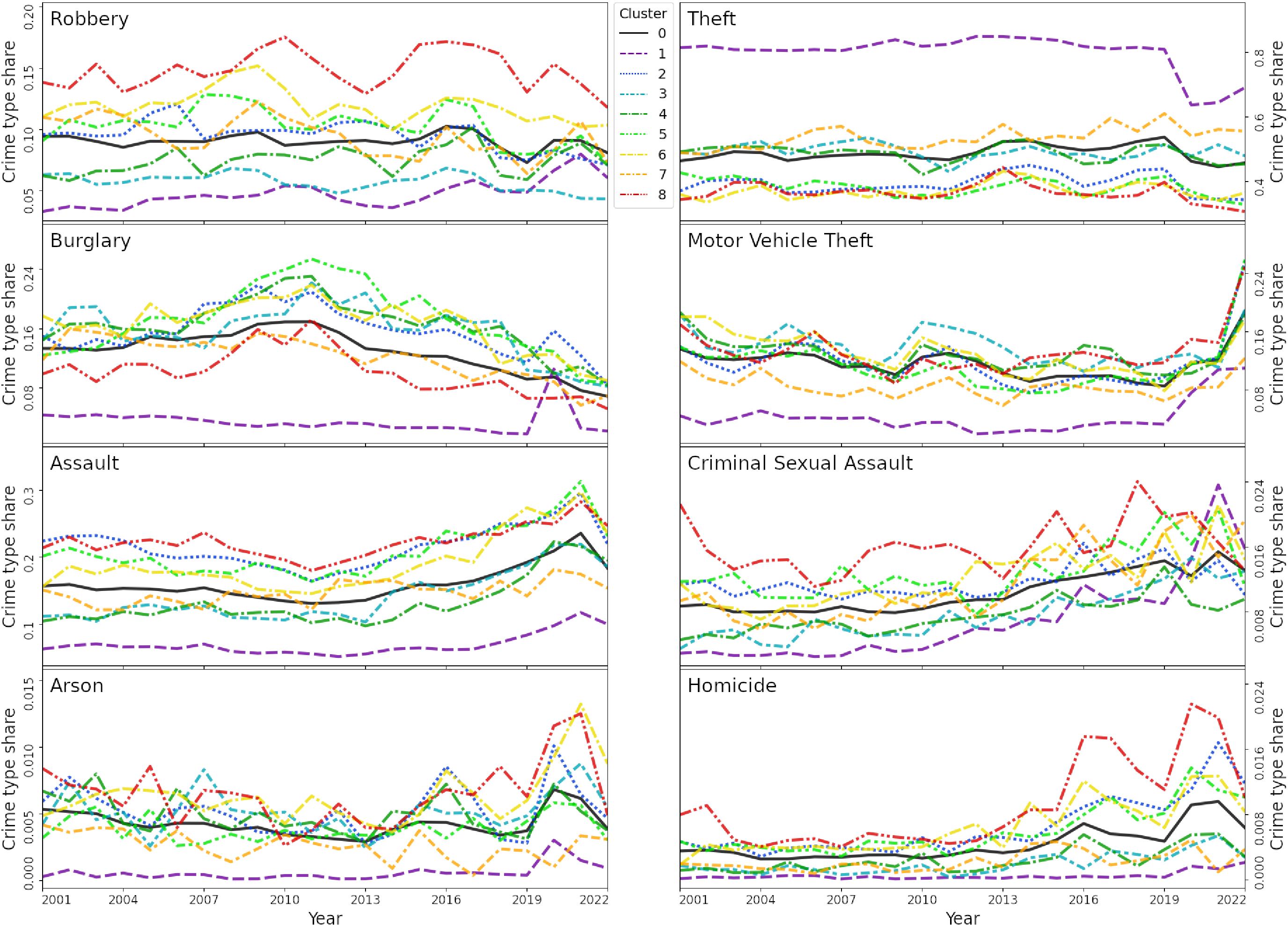}
\caption{Cluster composition for crime reports in the City of Chicago from 2001--2022. The eight panels for separate Part I crimes show the evolution of the share of each cluster's total incidents that a given crime type contributes to that cluster over the years. The average of non-clustered outliers is shown as a solid black line.}
\label{fig:figure_3}
\end{figure*}

To resolve these issues, we plot percentual shares per total cluster-associated incidents in Figure~\ref{fig:figure_3}, showing how different crime types contribute to a given cluster's composition, and to provide both the intra-cluster evolution of relative crime shares over time and an inter-cluster comparison. The plots also provide a confirmation that hot spots deviate from, and are centered on, the overall crime profile of surrounding non-clusterered data points. There are considerable deviations in cluster composition from the non-clustered average, as well as inter-cluster differences. An example is the already mentioned Chicago Loop area in Cluster 1, which differs from other clusters and outliers in multiple ways. This includes a much higher share of theft and stark differences in the evolution of burglary, with such differences playing an important part in policy planning.

Our analysis also provides a striking example of the COVID-19 pandemic's impact on both crime types, with an extreme drop and spike for theft and burglary, respectively. As the area is focused on entertainment and commercial venues, the lockdown measures following the pandemic decreased the number of potential targets for theft, whereas non-occupied commercial buildings in an overall emptier area provided more opportunities for break-ins.

Less obviously, the number of arson cases in Chicago, especially in Cluster 1, experienced a notable rise in the same year. Part of the reason for this development are the widespread protests and associated instances of arson following the killing of George Floyd instead of the pandemic, highlighting the risk of incorrect conclusions \citep{Reny2021}. Other Part I crimes, particularly assault, follow a much more consistent trajectory across clustered and non-clustered incident reports, only differing in the relative share, while homicide experiencing a growing spread from the early 2010s onward, especially for the Austin community area. 

\begin{table*}[!htb]
\caption{Total number of Part I crime incident reports for Clusters C1--C8 and non-clustered outliers in the City of Chicago from 2001--2022. The table lists the reports of robbery, theft, burglary, motor vehicle theft, assault, criminal sexual assault, arson, and homicide, with the first two letters of each crime type indicated in the column headers.}
\begin{footnotesize}
\begin{center}
\begin{tabular}{lllllllll}
\hline
Subset & RO & TH & BU & MO & AS & CR & AR & HO \\
\hline
Outliers & 205839 & 1101411 & 303089 & 266959 & 355058 & 22301 & 9523 & 9011 \\
C1 & 12105 & 210119 & 8792 & 10712 & 17839 & 1678 & 126 & 114 \\
C2 & 14677 & 59148 & 24423 & 17978 & 32007 & 1715 & 774 & 857 \\
C3 & 3552 & 30148 & 9903 & 8670 & 8012 & 417 & 309 & 100 \\
C4 & 7854 & 52255 & 18122 & 14092 & 13379 & 753 & 536 & 223 \\
C5 & 14648 & 53763 & 24318 & 16191 & 28340 & 1727 & 529 & 755 \\
C6 & 9903 & 30043 & 14007 & 11082 & 14945 & 927 & 496 & 449 \\
C7 & 6818 & 37554 & 8898 & 6103 & 10216 & 751 & 193 & 146 \\
C8 & 12811 & 31584 & 9031 & 11387 & 19130 & 1376 & 549 & 725 \\
\hline
\end{tabular}
\end{center}
\label{tab:table_3}
\end{footnotesize}
\end{table*}

Motor vehicle theft, on the other hand, shows a marked spike in Figure~\ref{fig:figure_1} for the city as a whole, but also across clusters in a sharp upward trend \citep{Lopez2021}. This offense is known to be a keystone crime that facilitates the commission of other types, and is also linked to the general uptick in violent crime rates after the onset of pandemic measures in Figure~\ref{fig:figure_2} \citep{Farrell2011}. As these are relative measures of crime type composition, allowing for a simultaneous depiction of crime type, cluster membership, and time, Table~\ref{tab:table_3} also provides total counts for reported incidents among clusters and outliers over the investigated time span.

\subsection{Distributions and clumping of crime}
\label{subsec:distributions_and_clumping_of_crime}

Lastly, we target the intra-cluster uniformity of incident reports, using the calculation of two-point autocorrelations described in Section~\ref{subsec:uniformity_and_autocorrelations} as a methodology import from cosmology into a domain application of operational research. We opt for the Landy-Szalay estimator from Equation~\ref{eq:2pCF_ls} due to its low sensitivity to the sample size and its ability to implement edge corrections. For a comparison of methods to compute the two-point autocorrelation function, as well as to the alternatives by \citet{Ripley1976}, see \citet{Kerscher2000}.

\begin{figure*}[!htb]
\includegraphics[width=\textwidth]{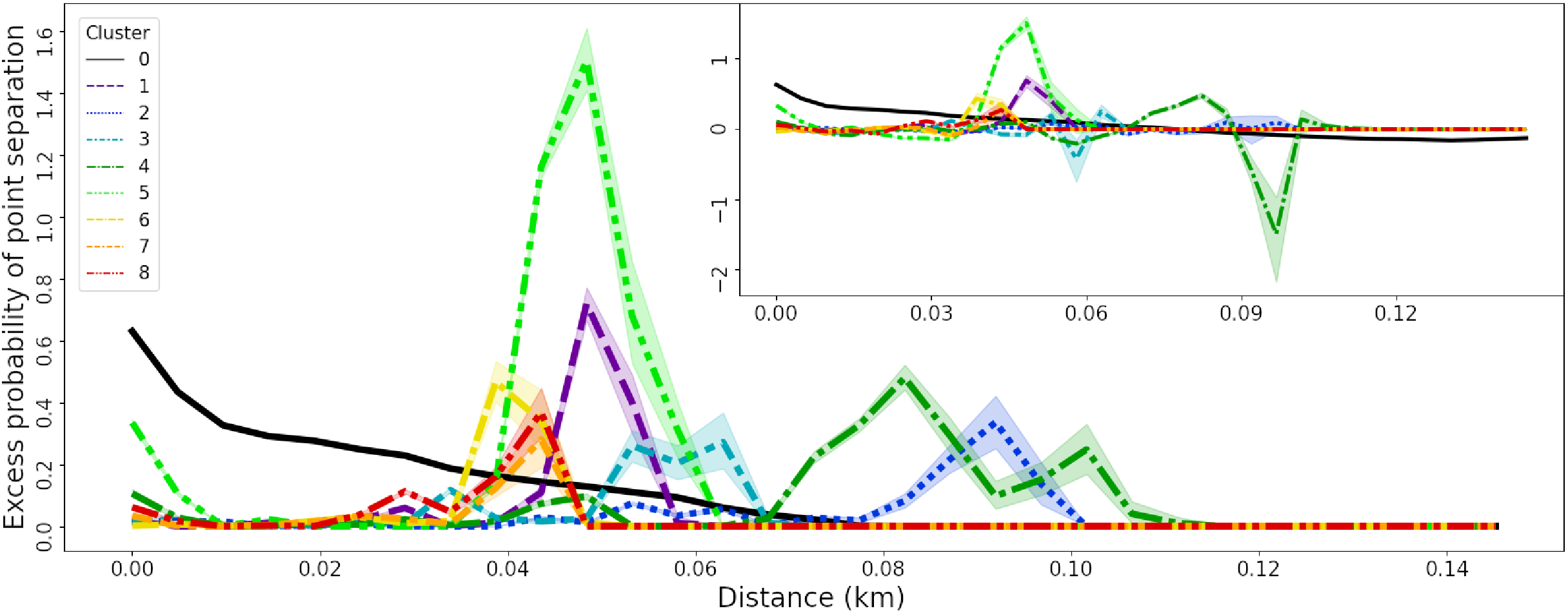}
\caption{Spatial two-point autocorrelations for identified clusters in Part I crime incidents for the City of Chicago from 2001-2022, with shaded regions indicating 95\% confidence intervals for multiple runs. The main panel shows correlations at different distance radii with a threshold of zero to alleviate the challenge of locations inaccessible for data collection, while the subplot depicts the non-thresholded correlations.}
\label{fig:figure_4}
\end{figure*}

The main panel of Figure~\ref{fig:figure_4} shows the function values at different distance radii, for which we use a zero-threshold to visualize positive excess probabilities without below-average probabilities induced by empty areas as described in Section~\ref{subsec:clustering_and_hot_spot_description}. We sample 10,000 data points per simulation, and run 100 simulations to create confidence intervals around the plotted averages per cluster, with measurements in ten-meter incremements to capture irregularities at different distances.  The surrounding outliers are defined as Cluster 0 and plotted as well to compare intra-cluster results to the non-clustered remainder of the city.

Intra-cluster deviations from uniform distributions beyond the non-clustered average are present for all hot spots, albeit at varying distances. The most salient clumping effect for crime can be seen in Cluster 5, which is in agreement with \citet{Schnell2017}, whose results demonstrate that violent crime is concentrated in specific areas in southern Chicago, with notable differences between more fine-grained adjacent neighborhood clusters.

The subplot of Figure~\ref{fig:figure_4}, in contrast, omits the zero-threshold, confirming the effect of empty spots due to block-level data as a data limitation. This is particularly noticeable for Clusters 3 and 4, which were already pointed out in Section~\ref{subsec:clustering_and_hot_spot_description} due to the presence of a large airport in Cluster 4 in particular, as well as multiple parks and cemeteries. These negative values as the result of empty flecks in hot spots also resemble the effects of cosmic voids in the source field of this method \citep[see, for example,][]{Inoue2007}. 

The results of this section demonstrate intra-cluster deviations from an uniform crime distribution when accounting for data limitations, highlighting the necessity to tailor the policing of hot spots through patrolling efforts to substructures within them. They also showcase said limitations due to the data collection process and modifications to allow for open-source dissemination. 

\section{Disussion and study limitations}
\label{sec:discussion_and_study_limitations}

The application of accurate estimates of the presence and evolution of crime, both spatially and over time, is a core concept in improving public safety measures through urban planning and methodologically sound analyses in operational research to inform modern policing approaches. Aside from direct applications to criminal justice, areas such as street environment design and transit planning due to travel modes being influenced by personal safety concerns rely on estimates of risk clusters \citep{Halat2015, Fabusuyi2018, Hong2019, Mao2021}.

Our work provides several new insights on spatio-temporal variations in the evolution of hot spot composition, the impact of the COVID-19 pandemic and the notable role of primary area functions in this process, the transfer of mathematical methods from other fields, and effective hot spot policing. As our study rescales distances based on point densities, we detect variable-density clusters to take the heterogeneous nature of urban population distributions into account. 

When compared to a non-adjusted cluster analysis, multiple areas remain unchanged, including the region of Gage Park and Chicago Lawn, South Shore and South Chicago, and the area around the Chicago Loop. Effects can also be observed in the extraction of more spread-out clusters. If, however, a specific prevention strategy targets high-density areas regardless of composition and density profiles, then a non-adjusted implementation can be more useful, and a compromise can be reached by combining results from both methods, as shown in Figure~\ref{fig:figure_2}.

Practitioners are often interested in the highest few percentiles in the density distribution of crime, which is bolstered by hot spots persisting for over a decade in 5\% of block-long street segments \citep{Weisburd2004, Braga2014}. Here, a thresholding method that horizontally slices the crime landscape based on a kernel density estimate of the underlying distribution, as done by \citet{Moews2021}, can also be applied to our approach. 

For follow-up research, we recommend the investigation of temporal cluster movements using, for example, Wasserstein distances, and collaborations with the public sector to introduce modern clustering methods to the toolbox of crime analysts. While the operational research literature contains a wealth of related research, including on improved policing, its use in practical settings is far from certain. One factor that affects these impact outcomes is rapport between researchers and practitioners, and \citet{Weisburd2005} document a strong relationship between an early adoption and interactions with the research community \citep{Newsome2008}.

Notably, \citet{Novak2016} report in their work with police officers that they were unable to implement a randomized study due to the department's reluctance to give up control over patrol selection for some areas in favor of a stronger research design. While there is no easy fix for these disagreements, their research suggests that frequent discourse with practitioners,through seminars or collaborative workshops, is a step in the right direction to boost methodological adoption. This is important due to discrepancies between officers' perceptions on where areas of serious crime are concentrated and the empirical evidence in their collected data, which can be influenced by the ethnic composition of neighborhoods \citep{Haining2007, Xie2012}.

As urban environmental circumstances change, so does the composition of crime. An initial drop in crime rates has been investigated early on in the COVID-19 pandemic, confirming the impact of lockdowns on routine activities on larger scales \citep{Stickle2020}. \citet{Schleimer2021} paint a more nuanced picture for urban environments in the United States and report an increase and decrease in property and violent crimes, respectively. They correlate adherence to lockdown measures to increases in arson, burglary, and motor vehicle theft, while other Part I crimes decrease, with the exception of homicide remaining indistinguishable from the null hypothesis. These findings are linked to the routine activities approach to crime, which addresses the change in such routines affecting the opportunities for different cime types \citep{Cohen1979}.

Our findings follow up on these works and showcase the impact of the pandemic and its associated lockdowns and social distancing measures for the City of Chicago. The role of specific hot spots in the city's life in particular is highlighted across our results, most notably through the respective drop and rise in theft and burglaries for the Chicago Loop area. As we investigate crime shares of these clusters, our experiments also demonstrate the changing risk of being exposed to different crime types relative to the overall crime rate in a given area, which is relevant for public safety considerations, as well as an increasing spread in homicide shares for hot spots over time. A natural explanation for the overall increase in criminal sexual assault in terms of its share of reported crime incidents is the varying difficulty to address crime types based on motivating factors. Sexual violence in particular is, much unlike property crimes such as burglary as well as planned or unplanned assault, inherently linked to power dynamics, and a successful reduction likely has to target patriarchal structures \citep{Patel2001, Canan2019}.

We reserve the rest of this section to the discussion of data bias as an area that is, aside from a mention of potential discrepancies between collected data and underlying distributions, often overlooked in related research. The above paragraph is linked to this topic, as reporting of criminal sexual assault is often lower than for other Part I crimes and suffered from a decrease during the COVID-19 pandemic due to a reluctance to visit hospitals \citep{Sorenson2021}. 

Just like perpetrator motivations, reasons for not reporting also differ from other violent crimes \citep{Thompson2007}. For crime reporting more broadly, police-community relations play a crucial role in the willingness to report incidents, and \citet{Kochel2018} concludes that residents' perceived police legitimacy has a marked influence on reporting, with a perception of the police as just and legitimate playing a larger role than notions of effectiveness \citep{Verschelde2012, Nepomuceno2022}. At the same time, biases in police action, as documented by \citet{Haining2007}, have the potential to further diminish perceived legitimacy.

\citet{Slocum2010} also document an inverse relationship of neighborhood poverty and crime reporting intentions, although a more in-depth recent analysis of the National Crime Victimization Survey shows that this effect is limited to male residents, highlighting the role of structural inequalities in a multifaceted environment \citep{Zaykowski2019}. Longitudinal studies of an earlier version of this data by \citet{Conaway1994} links reporting likelihoods to prior victimization experiences and police interactions, and related research demonstrates that crime reporting has increased \citep{Baumer2010}. Such trends also vary between U.S. metropolitan areas, which presents a complication for uniform collections of crime statistics \citep{Xie2014}.

Reporting biases due to this wide variety of factors, many of them hard to quantify and varying both between places and over time, result in data biases, which impacts the robustness of mathematical analyses in terms of generalizing from incident reports to true crime rates. The above paragraphs paint a picture of crime reporting dynamics that directly impact biases in datasets on reported incidents used in operational research. While we have to conclude that there is no simple solution for these kinds of biases due to the interconnected sociocultural factors involved, these limitations should be kept in mind for any area-specific crime analyses.

\section{Conclusions}
\label{sec:conclusion}

The analysis of crime patterns to inform policy decisions on crime prevention measures and urban planning, as well as policing efforts such as coverage priority and optimized patrolling, are an important application area of modern operational research. At the same time, adoption rates of mathematical methodology by practitioners in the criminal justice sector have been identified as a bottleneck, and easy-to-interpret visualizations and analyses play an crucial role in the applicability and interpretations of results to drive decision-making based on empirical information.

In this paper, we adapt and apply current advances in distance matrix rescaling and geospatial clustering, as well as a knowledge transfer from the field of cosmology, to the study of Part I crime types in operational research. Focusing on the City of Chicago due to the availability of large-scale complete and recent uniform data collections for, we provide the first application of continuous distance rescaling for the identification of structurally similar and arbitarily-shaped crime hot spots, as well as an analysis of their long-term evolution from 2001 to the end of 2022, and demonstrate urban feature effects on data reliability through spatial autocorrelations.

Our work provides several additional novel contributions to the literature, both for the operational research community active in this area of research and for practitioners and policy makers. Results on clumping effects offer insights into the spatial intra-cluster composition, which warrants more fine-grained policing approaches instead of currently widely used epicenter approaches subject to smoothing effects. We find these features to outshine the overall granularity observed in the remainder of the city outside of hot spots, and join existing calls regarding the relevance of substructures within hot spots for the development of effective crime prevention.

The findings highlight the impact of the recent COVID-19 pandemic through associated lockdowns and social contact avoidance measures, and one particularly important finding is the notable effect of primary city area functions on changing crime share compositions during this period. For follow-up research, we propose the analysis of historical analogues for expected impacts on crime for the optimization of police response planning. Here, our findings of the local dependence of such composition changes on area functions are especially important for appropriate capacity planning in practical scenarios to avoid inaccurate one-size-fits-all approaches to policing.

We also discuss the risk and contributing factors of data biases as a result of sociocultural influences, as the relationship of residents with local police forces and the biases among the former can negatively impact data collection efforts, which is an area often neglected in related research. Similarly, we discuss differences in crime suppression effects for various types of offenses, using the example of sexual violence in particular as a serious crime type that remains difficult to address due to underlying motivations. While the presented work is not focused on the field of community operational research, police-community relations also have a marked impact on the success of crime prevention measures. Our discussion thus motivates the relevance of these hard-to-quanitfy factors in the mathematical analysis of crime and subsequent interpretations.

\section*{Acknowledgments}

We would like to thank Antonia Gieschen for discussions on the underlying clustering approach, as well as Joe Zuntz for additional insights into the interpretation of two-point statistics in cosmology. We also wish to express our thanks to Jaime R. Argueta Jr., who initially recommended the Chicago Data Portal as a suitable source, and to the City of Chicago and the Chicago Police Department for their frequently updated and comprehensive publication of open-access data.

\section*{Declaration of interest statement}

Declarations of interest: None. This research did not receive any specific grant from funding agencies in the public, commercial, or not-for-profit sectors.

\bibliographystyle{apalike}
\bibliography{ref.bib}

\end{document}